\documentclass[12pt]{article}
\usepackage{amssymb,amsmath,epsfig}

\begin{document}

\title{\bf Statefinder Diagnostic for Dark Energy Models in Bianchi I Universe}
\author{M. Sharif \thanks {msharif.math@pu.edu.pk} and Rabia Saleem
\thanks{rabiasaleem1988@yahoo.com}\\
Department of Mathematics, University of the Punjab,\\
Quaid-e-Azam Campus, Lahore-54590, Pakistan.}

\date{}
\maketitle

\begin{abstract}
In this paper, we investigate the statefinder, the deceleration and
equation of state parameters when universe is composed of
generalized holographic dark energy or generalized Ricci dark energy
for Bianchi I universe model. These parameters are found for both
interacting as well as non-interacting scenarios of generalized
holographic or generalized Ricci dark energy with dark matter and
generalized Chaplygin gas. We explore these parameters graphically
for different situations. It is concluded that these models
represent accelerated expansion of the universe.
\end{abstract}

\section{Introduction}

Astronomical observations of Type Ia supernova \cite{1} indicate
that our universe is expanding with accelerating velocity rather
than slowing down. This cosmic expansion is confirmed by some
other independent observations like CMBR \cite{2}, SDSS \cite{3},
WMAP \cite{4} etc. An exotic form of matter with large negative
pressure is found to be responsible for this cosmic expansion
known as dark energy (DE) which occupies $2/3$ of the total energy
of our universe. All attempts indicate that nature as well as
origin of DE is still enigmatic and is a mystery for scientists.
The simplest candidate for DE is the cosmological constant with
equation of state (EoS) $\omega=-1$. The value of this
cosmological constant is very small as compared to quantum field
theory which is known as cosmological constant problem, i.e.,
fine-tuning and cosmic-coincidence problems. The nature of DE has
been explored by classifying the behavior of EoS parameter like
quintessence \cite{5}, k-essence \cite{6}, tachyon field \cite{7},
phantom model \cite{8} and Chaplygin gas \cite{9} etc. However,
none of these models is very successful.

Holographic principle \cite{10} is a good attempt in this
direction which explains some problems of cosmological constant
and DE. According to this principle, the degree of freedom in a
bounded system should be finite and it scales with its boundary
area but not with its volume. Cohen et al. \cite{11} proposed a
relationship between ultraviolet (UV) and infrared (IR) cut-offs
due to a limit set by the formation of a black hole, i.e.,
$L^3\rho_{\nu}\leq LM_{p}^{2}$. Here $\rho_{\nu}$ is the vacuum
energy density associated with UV cut-off, $L$ is the IR cut-off
and $M_P=(8\pi G)^{-\frac{1}{2}}$ is the reduced Planck mass.

In order to discuss an accelerated universe, three different values
of $L$ have been introduced, i.e., apparent, particle and event
horizon. It is found \cite{12} that the first two horizons do not
give accelerated expansion of the universe. The best choice is the
future event horizon for which $\omega<-\frac{1}{3}$, a sufficient
condition for cosmic acceleration. Zhang \cite{13} pointed out that
our universe is bounded by the future event horizon for which vacuum
energy density is converted to holographic DE (HDE) density, i.e.,
$\rho_\nu=3c^2M_p^{2}L^-2$, where $c$ is a dimensionless parameter.
This model has been used to investigate the cosmic evolution by
different people \cite{14}.

Holographic DE looks reasonable as it resolves some problems related
to DE but it also suffers with causality problem, i.e., future event
horizon is presumed in this model. Gao et al. \cite{15} gave the
idea that DE density and Ricci scalar are proportional to each
other, i.e., $\rho_x\propto R$ - known as Ricci dark energy (RDE).
This model of DE is phenomenological viable as it gives results
consistent with observational data. It also alleviates the causality
as well as cosmic coincidence problem. The Ricci scalar for flat
universe is $6(\dot{H}+2H^2)$ for which RDE density becomes
$\rho_r=3c^2(\dot{H}+2H^2)$. Chattopadhyay \cite{16} showed that
when the generalized RDE (GRDE) is considered in Horava-Lifshitz
gravity, it behaves like quintessence for $c^2>, =, <\frac{1}{2}$.

Feng and Li \cite{17} investigated viscous RDE model by assuming
that there is a bulk viscosity in the linear barotropic fluid and
RDE. The RDE model can be obtained by choosing the causal scale as
IR cut-off. Kim et al. \cite{18+} showed that the accelerating phase
of the RDE is that of a constant DE model for FRW universe. Recent
work in RDE and HDE models is demonstrated in \cite{19}. Xu et al.
\cite{21} proposed two models of DE, i.e., generalized HDE (GHDE)
and GRDE whose energy densities are
\begin{eqnarray*}
\rho_h=3c^2m_p^2H^2f(\frac{R}{H^2}),\quad \rho_r=3c^2m_p^2R
g(\frac{H^2}{R}),
\end{eqnarray*}
where c is a constant, $f(x)=\alpha x+(1-\alpha),~g(y)=\beta
y+(1-\beta),~(\alpha, \beta$ are constants) are positive functions.
We can also recover the original energy densities of GHDE and GRDE
by assuming $\alpha, \beta=0,1$. Notice that GHDE model is converted
into GRDE model by replacing $\alpha$ with $1-\beta$. In a recent
paper \cite{22}, the accelerating universe is investigated through
the deceleration and statefinder parameters for GHDE and GRDE models
by using FRW metric.

In this paper, we consider LRS Bianchi I (BI) universe model
composed of DM, generalized Chaplygin gas (GCG) and DE with
GHDE/GRDE model to discuss evolution of the universe. The plan of
the paper is as follows: In section \textbf{2}, the statefinder
parameters are calculated for the combination of two fluids. Section
\textbf{3} is devoted to explore the EoS, the deceleration
parameters and statefinder diagnostic pair for GHDE/GRDE model
without DM. We investigate these parameters with DM in
non-interacting and interacting scenarios in section \textbf{4}. In
section \textbf{5}, these parameters are discussed when the universe
is composed of GHDE/GRDE with GCG in non-interacting and interacting
scenarios. In the last section, we summarize the results.

\section{Statefinders for Two Fluid System}

Our universe appears homogeneous and isotropic on large scale at the
present time. The existence of the anisotropy at early times is a
natural phenomenon. We observe the anisotropy in galaxies, clusters
and super clusters today. It would be appropriate to discuss a
geometry that is more general than the isotropic and homogeneous FRW
geometry. A Bianchi type I model being the straightforward
generalization of the flat FRW model, is one of the simplest models
of the anisotropic universe. This model describes a homogeneous,
spatially flat and anisotropic universe.

In this section, we formulate the field equations, deceleration
parameter and statefinder parameters for LRS BI universe model. We
assume that the fluid is a combination of DM and DE. The line
element of BI model is given as follows
\begin{equation*}
ds^2=-dt^2+A^2(t)dx^2+B^2(t)(dy^2+dz^2),
\end{equation*}
where $A$ and $B$ are scale factors. We use the well-known
condition $A=B^m$ \cite{23}, where $m\neq1$ is a positive
constant. Consequently, the above metric reduces to
\begin{equation}\label{2}
ds^2=-dt^2+B^{2m}(t)dx^2+B^2(t)(dy^2+dz^2).
\end{equation}
The field equations corresponding to perfect fluid turn out
to be
\begin{eqnarray}\label{3}
(2m+1)\frac{\dot{B}^2}{B^2}=8\pi \rho,\\\label{4}
2\frac{\ddot{B}}{B}+\frac{\dot{B}^2}{B^2}=-8\pi p,\\\label{5}
m^2\frac{\dot{B}^2}{B^2}+(m+1)\frac{\ddot{B}}{B}=-8\pi p.
\end{eqnarray}
Equation (\ref{3}) can be written as
\begin{equation}\label{6}
H_2^{2}=\frac{1}{1+2m}(\rho_m+\rho_X),\quad H_2=\frac{\dot{B}}{B}.
\end{equation}
where $H_2$ is the directional Hubble parameter while $\rho_m$ and
$p_m$ are the energy density and pressure of matter respectively,
$\rho_X=\rho_h,~p_X=p_h$ are the energy density and pressure for
GHDE and $\rho_X=\rho_r,~p_X=p_r$ are the energy density and
pressure for GRDE, respectively. The conservation equation yields
\begin{equation}\label{7}
\dot{\rho}_m+\dot{\rho}_X+(m+2)(\rho_m+\rho_X+p_m+p_X)H_2=0.
\end{equation}
Taking derivative of Eq.(\ref{6}) and using (\ref{7}), we obtain
\begin{equation}\label{8}
\dot{H}_2=-\frac{(m+2)}{2(1+2m)}(\rho_m+\rho_X+p_m+p_X).
\end{equation}

Sahni et al. \cite{24} introduced a new dimensionless static,
statefinder, which can differentiate between different types of DE
models and might be a good diagnostic of cosmological models. As the
statefinder diagnostic pair depends upon the scale factor, so we can
say that this pair is a geometrical diagnostic in the sense that it
is constructed from a spacetime metric directly. This pair examines
expansion of the universe at large scale as it involves third
derivative of the scale factor $B(t)$.

The deceleration parameter, $q$, and statefinder diagnostic pair
$\{r,s\}$ in terms of scale factor are formulated as follows
\begin{eqnarray}\nonumber
q&=&-\left[\frac{(m-1)}{(m+2)}+\frac{3B\ddot{B}}{(m+2)\dot{B}^2}
\right],\\\nonumber
r&=&\frac{(m-1)(m-4)}{(m+2)^2}+\frac{9(m-1)B\ddot{B}}
{(m+2)^2\dot{B}^2}+\frac{(m+2)B^2\dddot{B}}{3\dot{B}^3},\\
s&=&\frac{r-1}{3(q-\frac{1}{2})}\nonumber.
\end{eqnarray}
These can be expressed in the form of pressure and density as
\begin{eqnarray}
q&=&\frac{1}{2}+\frac{3}{2}\left(\frac{p_m+p_X}{\rho_m+\rho_X}
\right)\label{8+},\\\nonumber
r&=&\left[\frac{3m^5+21m^4+54m^3-3m^2+15m+72}{18(m+2)^2}\right]\\\nonumber
&+&\left[\frac{-9(m-1)}{2(m+2)}+\frac{(m-1)
(m+2)^2}{9}+\frac{(m+2)^3}{18}-\frac{(m+2)}{2}\right]\left
(\frac{p_m+p_X}{\rho_m+\rho_X}\right)\nonumber\\
&+&\frac{(m+2)^3}{6(\rho_m+\rho_X)}\left
[\frac{\partial p_m}{\partial\rho_m}(p_m+\rho_m) +\frac{\partial
p_X}{\partial\rho_X}(p_X+\rho_X)\right]\label{9}, \\\label{10}
s&=&\left[\frac{-(m-1)}{(m+2)}+\frac{2(m-1)(m+2)^2}{81}+\frac
{(m+2)^3}{81}-\frac{(m+2)}{9}\right]\nonumber\\&+&\left
[\frac{3m^5+21m^4+54m^3-21m^2-57m}{81(m+2)^2}\right]\left
(\frac{\rho_m+\rho_X}{p_m+p_X}\right)\nonumber\\&+&\frac{(m+2)^3}
{27(p_m+p_X)}\left[\frac{\partial
p_m}{\partial\rho_m}(p_m+\rho_m)+\frac{\partial
p_X}{\partial\rho_X}(p_X+\rho_X)\right].
\end{eqnarray}

\section{GHDE Model without Dark Matter}

Here we evaluate the EoS, the deceleration and statefinder
parameters when the universe is filled with GHDE/GRDE only. Equation
(\ref{6}) can be written as
\begin{equation}\label{11}
H_2^{2}=\frac{1}{1+2m}\rho_h.
\end{equation}
The Ricci scalar is given by
\begin{equation}\label{12}
R=-2\left[(m^2+2m+3)H_{2}^{2}+(m+2)\dot{H}_2\right].
\end{equation}
Using this value of $R$, we get GHDE density as
\begin{equation}\label{13}
\rho_h=\frac{c^2}{3}\left[-18(m+2)\alpha\dot{H}_2
+[(m+2)^2-(19m^2+40m+58)\alpha]H_{2}^{2}\right].
\end{equation}
Inserting this value of $\rho_h$ in Eq.(\ref{11}), we obtain
\begin{eqnarray}\label{14}
\dot{H}_2+\frac{(1+2m)H_{2}^{2}}{6\alpha
c^2(m+2)}[1-\frac{c^2}{3(1+2m)}[(m+2)^2-(19m^2+40m+58)\alpha]]=0
\end{eqnarray}
whose solution is
\begin{equation*}\label{15}
H_{2}^{2}=H_0^2B^{\mu},
\end{equation*}
where $H_0^2$ is an integration constant and
\begin{equation*}
\mu=-\frac{1+2m}{6(m+2)c^2\alpha}\left[1-\frac{c^2}
{3(1+2m)}[(m+2)^2- (19m^2+40m+58)\alpha]\right].
\end{equation*}
Substituting this value of $H_2^2$ in Eq.(\ref{11}), we have
\begin{equation}\label{16}
\rho_h=(1+2m)H_0^2B^{\mu}.
\end{equation}
Consequently, Eq.(\ref{8}) will become
\begin{equation}\label{17}
\dot{H}_2=-\frac{(m+2)}{2(1+2m)}(\rho_h+p_h).
\end{equation}
Inserting the values of $\dot{H}_2$ and $\rho_h$ from Eqs.(\ref{14})
and (\ref{16}), respectively in (\ref{17}), the pressure of GHDE can
be expressed as
\begin{equation}\label{18}
p_h=-(1+2m)\left[\frac{\mu}{(m+2)}+1\right]H_0^2B^{\mu}.
\end{equation}

The EoS parameter for GHDE is defined as
\begin{equation}\label{19}
\omega_h=\frac{p_h}{\rho_h}=-\left[\frac{\mu}{(m+2)}+1\right],
\end{equation}
which must satisfy $\omega_h<-\frac{1}{3}$ to show the expanding
universe. Under this condition, we get
$\alpha<\frac{(m+2)^2c^2-3(1+2m)}{(7m^2-8m+10)c^2}$ and the model
generates DE. The EoS parameter for GHDE is plotted against
$\alpha$ and $c$ shown in Figure \textbf{1}. This shows that
$\omega_h$ decreases from positive to negative as $\alpha$
decreases and $c$ increases. Equations (\ref{16}) and (\ref{18})
lead to
\begin{equation}\label{20}
\frac{\partial p_h}{\partial\rho_h}=\frac{\partial p_h/\partial B
}{\partial\rho_h/\partial B}=\omega_h.
\end{equation}
\begin{figure}
\center\epsfig{file=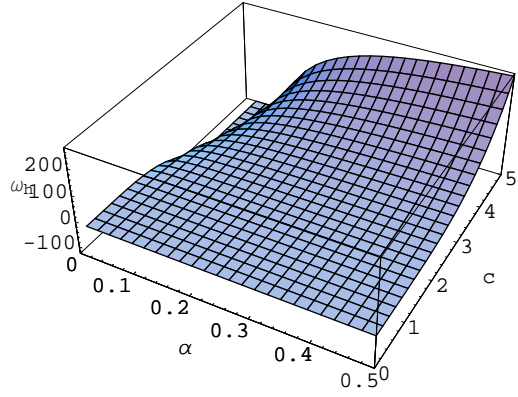, width=0.5\linewidth}\caption{Variation of
$\omega_h$ against $\alpha$ and $c$.}
\end{figure}
\begin{figure}
\center\epsfig{file=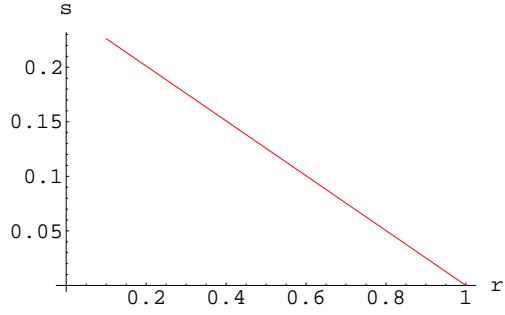, width=0.5\linewidth}\caption{Variation
of $r$ against $s$.}
\end{figure}
\begin{figure}
\center\epsfig{file=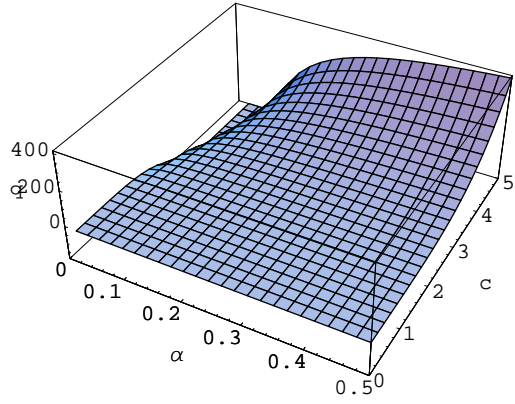, width=0.5\linewidth}\caption{Variation
of $q$ against $\alpha$ and $c$.}
\end{figure}
Using Eqs.(\ref{8}), (\ref{19}) and (\ref{20}) in
(\ref{8+})-(\ref{10}), we obtain the deceleration parameter and
statefinder diagnostics
\begin{eqnarray*}\label{21}
q&=&-1-\frac{3}{2}\frac{\mu}{(m+2)},\\\nonumber
r&=&\left[\frac{3m^5+21m^4+54m^3-3m^2+15m+72}{18(m+2)^2}\right]\nonumber\\
&-&\left[\frac{-9(m-1)}{2(m+2)}+\frac{(m-1)(m+2)^2}{9}
+\frac{(m+2)^3}{18}-\frac{(m+2)}{2}\right]\nonumber\\
&\times&\left[\frac{\mu}{(m+2)}+1\right]
+\frac{(m+2)^2\mu}{6}\left[\frac{\mu}{(m+2)}+1\right],\label{22}\\
s&=&\left[\frac{-(m-1)}{(m+2)}+\frac{2(m-1)(m+2)^2}{81}
+\frac{(m+2)^3}{81}-\frac{(m+2)}{9}\right]\nonumber\\
&-&\left[\frac{3m^5+21m^4+54m^3-21m^2-57m}{81(m+2)}\right]
\left[\frac{1}{\mu+(m+2)}\right]\nonumber\\
&-&\frac{(m+2)^2\mu}{27}\label{23}.
\end{eqnarray*}
The universe will be accelerating if $q<0$, i.e., when
$\alpha<\frac{(m+2)^2c^2-3(1+2m)}{(7m^2-8m+10)c^2}$. Figure
\textbf{2} shows that $s$ decreases as $r$ increases. The
deceleration parameter is shown in Figure \textbf{3} which indicates
that $q$ decreases from positive to negative values as $\alpha$
decreases and $c$ increases. If we replace $\alpha$ by $1-\beta$,
then all the above solutions are valid for GRDE. In this case,
$\beta>\frac{(6m^2-12m+6)c^2+3(1+2m)}{(7m^2-8m+10)c^2}$ for the
accelerating universe.

\section{GHDE Model with Dark Matter}

In this section, we evaluate the above mentioned parameters for
non-interacting and interacting scenarios when the universe is a
combination of GHDE and DM.

\subsection{Non-Interacting Case}

Here, we assume that the universe is filled with GHDE and
non-interacting DM. Consequently, Eq.(\ref{6}) yields
\begin{equation}\label{24}
H_2^{2}=\frac{1}{1+2m}(\rho_h+\rho_m).
\end{equation}
As there is no interaction between GHDE and DM, so these are
independently conserved. Equation (\ref{7}) leads to conservation
equations for DM and GHDE as follows
\begin{equation}\label{25}
\dot{\rho}_m+(m+2)(\rho_m+p_m)H_2=0,\quad
\dot{\rho}_h+(m+2)(\rho_h+p_h)H_2=0.
\end{equation}
For EoS $p_m=\omega_m\rho_m$, first of the above equation yields
\begin{equation}\label{27}
\rho_m=\rho_{m_0}B^{-(m+2)(1+\omega_m)},
\end{equation}
where $\rho_{m_0}$ is the constant of integration. Inserting the
values of $\rho_h$ and $\rho_m$ from Eqs.(\ref{13}) and (\ref{27})
in Eq.(\ref{24}), we obtain
\begin{equation*}\label{28}
\frac{d{H}_2^2}{dB}-\frac{\mu}{B}H_2^{2}=\frac{\rho_{m_0}B^
{-(m+2)(1+\omega_m)}}{6\alpha c^2(m+2) B}
\end{equation*}
whose solution is
\begin{equation}\label{29}
H_2^{2}=\frac{\rho_{m_0}B^{-(m+2)(1+\omega_m)}}{-6\alpha
c^2(m+2)^2\left[[\frac{\mu}{(m+2)}+1]+\omega_m\right]}+H_1^2
B^\mu,
\end{equation}
where $H_1$ is an integrating constant. Combining Eqs.(\ref{24}),
(\ref{25})-(\ref{29}), we obtain density and pressure
\begin{eqnarray}\label{30}
\rho_h&=&\nu\rho_{m_0}B^{-(m+2)(1+\omega_m)}+(1+2m)H_1^2B^\mu,\\\label{31}
p_h&=&\nu\rho_{m_0}\omega_mB^{-(m+2)(1+\omega_m)}
-(1+2m)[\frac{\mu}{(m+2)} +1]H_1^2B^\mu,
\end{eqnarray}
where
\begin{equation*}
\nu=\left[\frac{(1+2m)}{-6\alpha
c^2(m+2)^2\left[[\frac{\mu}{(m+2)}+1]+\omega_m\right]}-1\right].
\end{equation*}
\begin{figure}
\center\epsfig{file=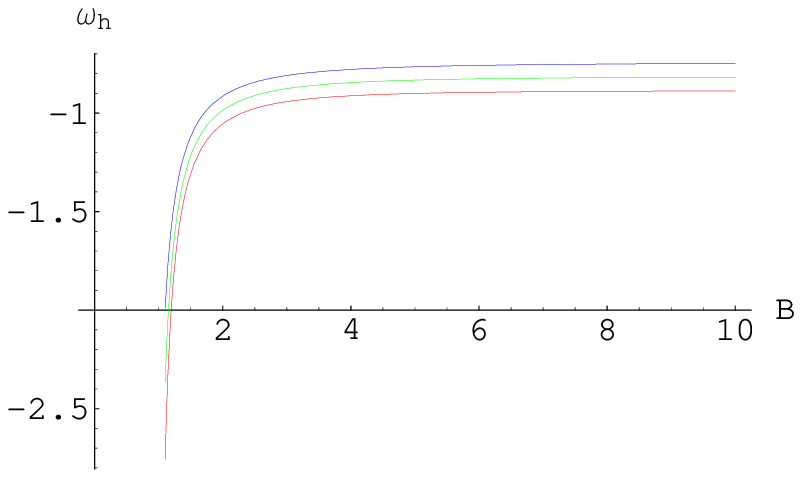, width=0.5\linewidth}\caption{Variation
of $\omega_h$ against $B$ for $\omega_m=0.01, ~\rho_{m_0}=1,
~H_1=1, ~c=2,$ and $\alpha=0.1, 0.12, 0.15.$}
\end{figure}
\begin{figure}
\center\epsfig{file=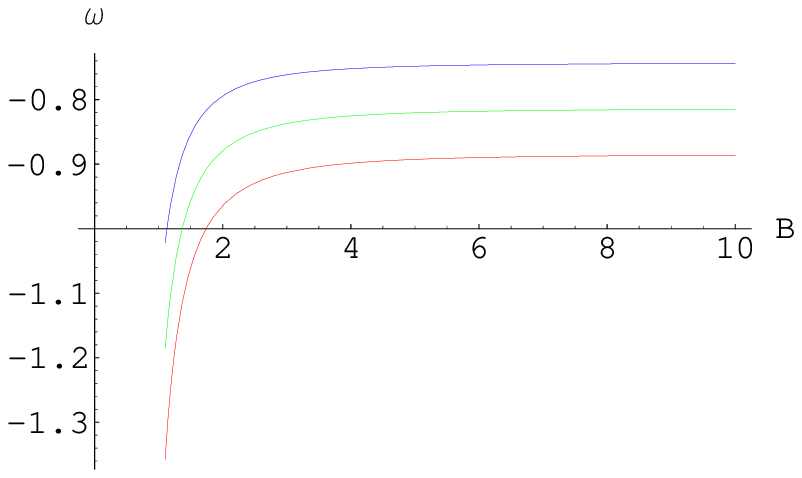, width=0.5\linewidth}\caption{Variation
of $\omega$ against $B$ for $\omega_m=0.01, ~\rho_{m_0}=1, ~H_1=1,
~c=2,$ and $\alpha=0.1, 0.12, 0.15.$}
\end{figure}
\begin{figure}
\center\epsfig{file=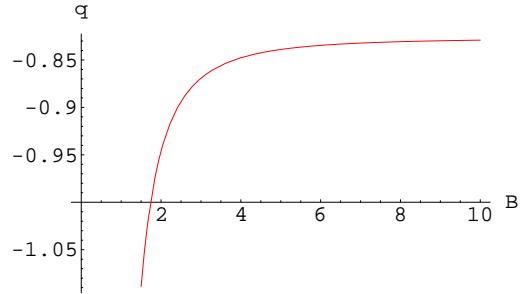, width=0.5\linewidth}\caption{Variation
of $q$ against $B$ for $\omega_m=0.01, ~\rho_{m_0}=1, ~H_1=1,
~c=2,$ and $\alpha=0.1.$}
\end{figure}
\begin{figure}
\center\epsfig{file=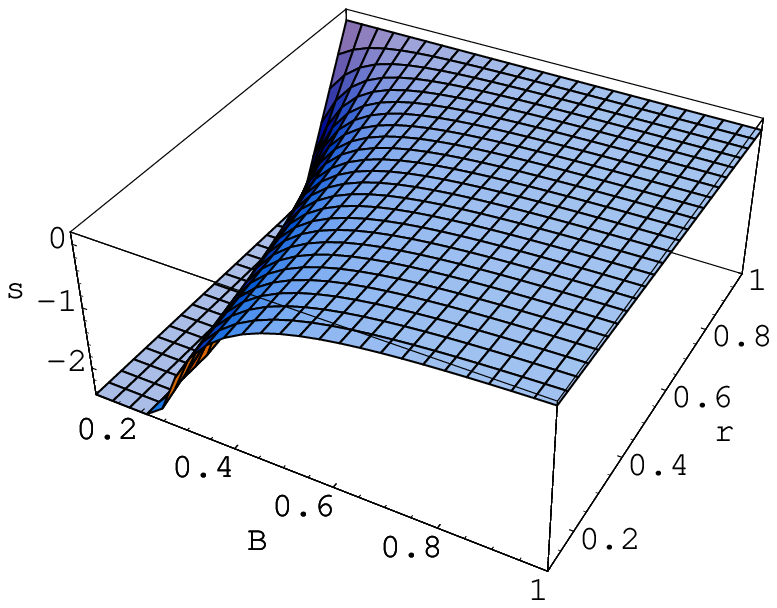, width=0.5\linewidth}\caption{Variation
of $s$ against $r$ and $B$ for $\omega_m=0.01, ~\rho_{m_0}=1,
~H_1=1, ~c=2$ and $\alpha=0.1.$}
\end{figure}
The EoS for GHDE and for the combined fluid is
\begin{eqnarray}\label{32}
\omega_h&=&\frac{p_h}{\rho_h}=\frac{\nu\rho_{m_0}\omega_mB^
{-(m+2)(1+\omega_m)}-(1+2m)[\frac{\mu}{(m+2)}+1]H_1^2B^\mu}
{\nu\rho_{m_0}
B^{-(m+2)(1+\omega_m)}+(1+2m)H_1^2B^\mu},\\\label{33}
\omega&=&\frac{p_h+p_m}{\rho_h+\rho_m}=\frac{1}{(\nu+1)\rho_{m_0}B^
{-(m+2)(1+\omega_m)}+(1+2m)H_1^2B^\mu}\nonumber\\
&\times&(\nu+1)
\rho_{m_0}\omega_mB^{-(m+2)(1+\omega_m)}-(1+2m)\nonumber\\&
\times&[\frac{\mu}{(m+2)}+1]H_1^2B^\mu.
\end{eqnarray}
We plot graphs of $\omega_h$ and $\omega$ against the scale factor
$B$ in Figures \textbf{4} and \textbf{5}, respectively for different
values of $\alpha$, which represent the evolution of the universe.
Also, the graphs of $q$ against $B$ as well as $s$ against $r$ and
scale factor $B$ are shown in Figures \textbf{6} and \textbf{7},
respectively. The deceleration parameter also generates negative
sign and represents expansion of the universe. Figure \textbf{7}
indicates that $s$ increases from negative to positive values as $r$
decreases and $B$ increases. We conclude that the non-interacting
case yields DE.

\subsection{Interacting Case}

Now we assume that the universe is a mixture of GHDE and DM
interacting with each other. The conservation equation (\ref{7})
takes the form
\begin{eqnarray}\label{34}
\dot{\rho}_m+(m+2)H_2(\rho_m+p_m)&=&-(m+2)\delta
H_2\rho_m,\\\label{35}
\dot{\rho}_h+(m+2)H_2(\rho_h+p_h)&=&(m+2)\delta H_2\rho_m,
\end{eqnarray}
where $(m+2)\delta H_2\rho_m$ is interaction, $\delta$ is known as
interaction parameter. Solving Eq.(\ref{34}), we obtain matter
density as
\begin{equation}\label{36}
\rho_m =\rho_{m_1}B^{-(m+2)(1+\omega_m+\delta)},
\end{equation}
where $\rho_{m_1}$ is the constant of integration. Equations
(\ref{13}), (\ref{24}) and (\ref{36}) yield the following solution
\begin{equation}\label{37}
H_2^{2}=\frac{\rho_{m_1}B^{-(m+2)(1+\omega_m+\delta)}}{-6\alpha
c^2(m+2)^2\left[[\frac{\mu}{(m+2)}+1]+(\omega_m+\delta)\right]}
+(H_{2}')^2B^\mu,
\end{equation}
where $H_{2}'$ is another integration constant. The corresponding
value of energy density and pressure can be obtained by combining
Eqs.(\ref{24}) and (\ref{35})-(\ref{37}) as follows
\begin{eqnarray}\label{38}
\rho_h&=&\gamma_0\rho_{m_1}B^{-(m+2)(1+\omega_m+\delta)}
+(1+2m)(H_{2}')^2B^\mu,\\\label{39}
p_h&=&\gamma_0\rho_{m_1}\omega_mB^
{-(m+2)(1+\omega_m+\delta)}-(1+2m)[\frac{\mu}{(m+2)}+1](H_{2}')^2B^\mu,
\end{eqnarray}
where
\begin{equation*}
\gamma_0=\frac{(1+2m)+6\alpha
c^2(m+2)^2\left[[\frac{\mu}{(m+2)}+1]+(\omega_m+\delta)\right]}
{-6\alpha
c^2(m+2)^2\left[[\frac{\mu}{(m+2)}+1]+(\omega_m+\delta)\right]}.
\end{equation*}
\begin{figure}
\center\epsfig{file=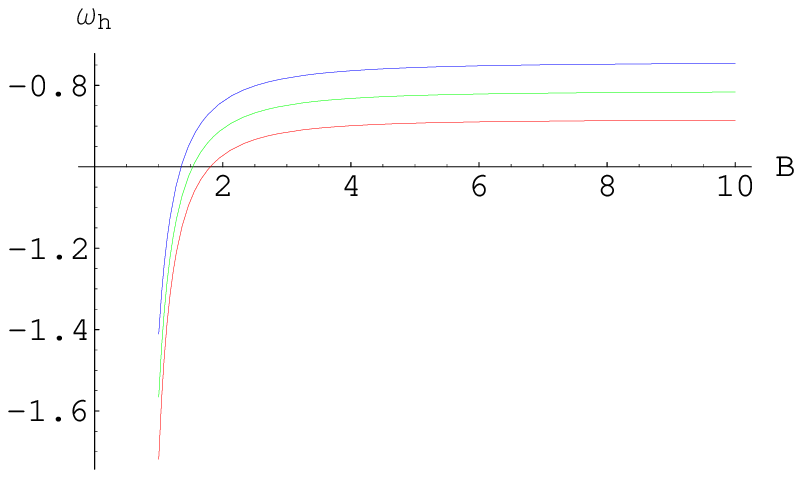, width=0.5\linewidth}\caption{Variation of
$\omega_h$ against $B$ for $\omega_m=0.01, ~\rho_{m_1}=1, ~H_{2}'=1,
~c=2, ~\delta=0.01,$ and $\alpha=0.1, 0.12, 0.15.$}
\end{figure}
\begin{figure}
\center\epsfig{file=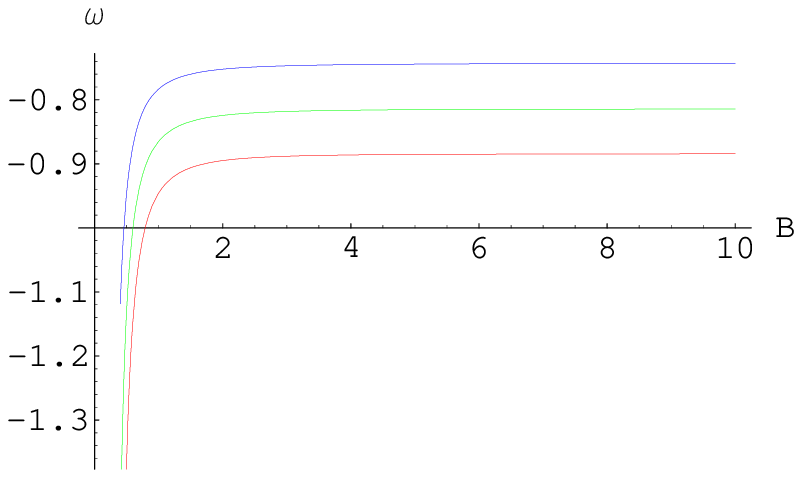, width=0.5\linewidth}\caption{Variation of
$\omega$ against $B$ for $\omega_m=0.01, ~\rho_{m_1}=1, ~H_{2}'=1,
~c=2, ~\delta=0.01,$ and $\alpha=0.1, 0.12, 0.15.$}
\end{figure}
\begin{figure}
\center\epsfig{file=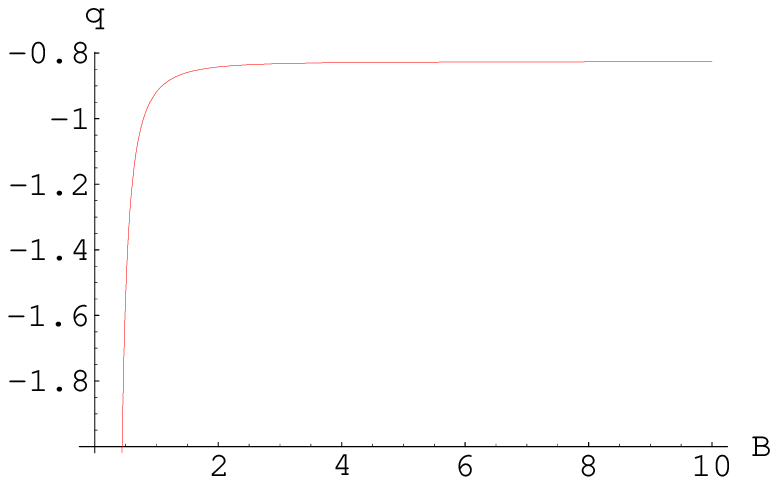, width=0.5\linewidth}\caption{Variation of
$q$ against $B$ for $\omega_m=0.01, ~\rho_{m_1}=1, ~H_{2}'=1, ~c=2,
~\delta=0.01, ~\alpha=0.1.$}
\end{figure}
\begin{figure}
\center\epsfig{file=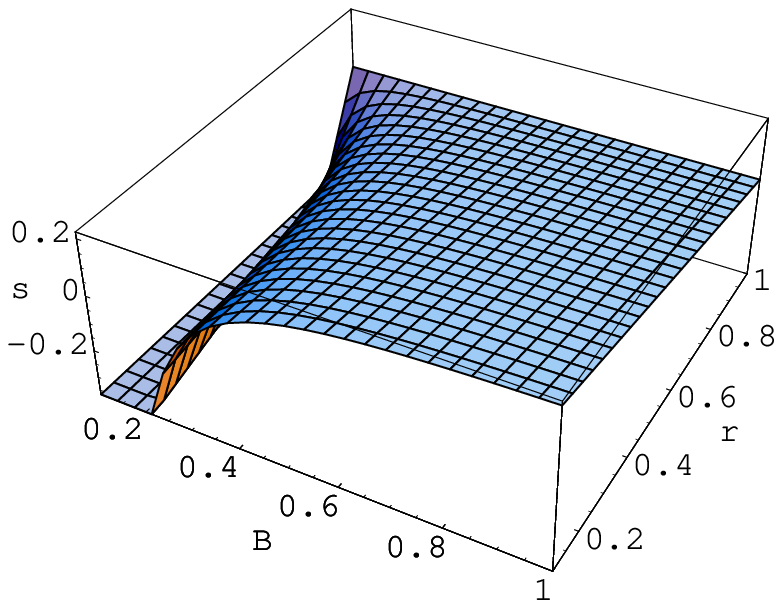, width=0.5\linewidth}\caption{Variation of
$s$ against $r$ and $B$ for $\omega_m=0.01, ~\rho_{m_1}=1,
~H_{2}'=1, ~c=2, ~\delta=0.01, ~\alpha=0.1.$}
\end{figure}
The EoS for GHDE can be expressed as
\begin{equation}\label{40}
\omega_h=\frac{\gamma_0\rho_{m_1}(\omega_m+\delta)
B^{-(m+2)(1+\omega_m+\delta)}-(1+2m)[\frac{\mu}{(m+2)}+1](H_{2}')^2B^\mu}
{\gamma_0\rho_{m_1}B^{-(m+2)(1+\omega_m+\delta)}+(1+2m)(H_{2}')^2B^\mu},
\end{equation}
and for interacting two fluid system, it takes the form
\begin{eqnarray}\label{41}
\omega&=&\frac{1}{(\gamma_0+1)\rho_{m_1}B^{-(m+2)(1+\omega_m+\delta)}
+(1+2m)(H_{2}')^2B^\mu}\nonumber\\
&\times&(\gamma_0+1)\rho_{m_1}(\omega_m+\delta)B^{-(m+2)
(1+\omega_m+\delta)}-(1+2m)(H_{2}')^2B^\mu\nonumber\\
&\times&[\frac{\mu}{(m+2)}+1].
\end{eqnarray}
The graphs \textbf{(8)-(11)} indicate that the interacting case also
provides the accelerated universe as all the above mentioned
parameters are negative and generates DE.

\section{GHDE Model with Generalized Chaplygin Gas}

The GCG is a route of investigation in which DM and DE are described
within one component model. This model predicts small scale
instabilities and oscillations at the perturbation level. The GCG
behaves like dust matter at early times and behaves like a
cosmological constant at late times. In \cite{25}, a correspondence
between HDE and CG has been established and showed that HDE could be
described by a scalar field in a certain way. The GCG has
interesting features like it has negative pressure and explains the
transition of our universe from decelerating to accelerating phase.
Keeping this motivation in mind, we consider generalized HDE and GCG
for interacting and non-interacting scenarios.

\subsection{Non-Interacting Case}

First we take the universe which is filled with GHDE and GCG in the
non-interacting case. The GCG is a perfect fluid given by \cite{26}
\begin{equation}\label{42}
p_c =\frac{-A}{\rho_c^{\gamma}},\quad A>0,\quad 0\leq\gamma\leq1.
\end{equation}
Equation (\ref{6}) can be written in the following form
\begin{equation}\label{43}
H_2^{2}=\frac{1}{1+2m}(\rho_h+\rho_c).
\end{equation}
In this case, the conservation equation (\ref{7}) for GHDE and GCG
become
\begin{equation}\label{44}
\dot{\rho}_c+(m+2)(\rho_c+p_c)H_2=0,\quad
\dot{\rho}_h+(m+2)(\rho_h+p_h)H_2=0.
\end{equation}
Solving the first of Eq.(\ref{44}), we get energy density of GCG
\begin{equation}\label{46}
\rho_c=[A+\rho_{c_0}B^{-(m+2)(1+\gamma)}]^{\frac{1}{1+\gamma}},
\end{equation}
where $\rho_{c_0}$ is an integration constant. Equation (\ref{42})
leads to pressure of GCG
\begin{eqnarray}\nonumber
p_c&=&\rho_{c_0}B^{-(m+2)(1+\gamma)}[A+\rho_{c_0}B^{-(m+2)
(1+\gamma)}]^{\frac{-\gamma}{1+\gamma}}\nonumber\\
&-&[A+\rho_{c_0}B^{-(m+2)(1+\gamma)}]^{\frac{1}{1+\gamma}}\label{47}.
\end{eqnarray}
Inserting Eqs.(\ref{46}) and (\ref{47}) in (\ref{43}), we obtain
\begin{equation}\label{49}
H_2^{2}=\frac{1}{6\alpha c^2(m+2)}e^{\mu x}
\int[A+\rho_{c_0}e^{-(m+2)(1+\gamma)x}]^{\frac{1}{1+\gamma}}e^{-\mu
x}dx+H_{c_0}^2B^\mu,
\end{equation}
where $x=\ln B$ and $H_{c_0}$ is another integration constant. The
corresponding GHDE density is obtained from Eqs.(\ref{43}),
(\ref{44}) and (\ref{49}) as
\begin{equation}\label{50}
\rho_h=\frac{(1+2m)}{6\alpha c^2(m+2)}e^{\mu x}I(x)
-[A+\rho_{c_0}B^{-(m+2)(1+\gamma)}]^{\frac{1}{1+\gamma}}
+(1+2m)H_{c_0}^2B^\mu,
\end{equation}
where
\begin{equation*}
I(x)=\int[A+\rho_{c_0}e^{-(m+2)(1+\gamma)x}]^{\frac{1}{1+\gamma}}
e^{-\mu x}dx.
\end{equation*}
The conservation equation of GHDE yields
\begin{equation}\label{51}
p_h=-\rho_h-\frac{1}{(m+2)}\frac{\partial\rho_h} {\partial x}.
\end{equation}
Equations (\ref{50}) and (\ref{51}) lead to
\begin{equation}\label{52}
p_h=\frac{(1+2m)}{6\alpha c^2(m+2)}e^{\mu x}I(x)
-[A+\rho_{c_0}B^{-(m+2)(1+\gamma)}]^{\frac{1}{1+\gamma}}
+(1+2m)H_{c_0}^2B^\mu.
\end{equation}

The EoS for GHDE is
\begin{eqnarray}\nonumber
\omega_h&=&(1+2m)[\frac{I(x)}{6\alpha
c^2(m+2)}+H_{c_0}^2]y(x)-[A+\rho_{c_0}B^{-(m+2)(1+\gamma)}]
^{-\frac{1}{1+\gamma}}\nonumber\\
&\times&[-\frac{(1+2m)}{6\alpha
c^2(m+2)^2}[A+\rho_{c_0}B^{-(m+2)(1+\gamma)}]^{\frac{1}{1+\gamma}}
-(1+2m)[\frac{\mu}{(m+2)}+1]\nonumber\\&\times&[\frac{I(x)}{6\alpha
c^2(m+2)}+H_{c_0}^2]y(x)+A[A+\rho_{c_0}B^{-(m+2)
(1+\gamma)}]^{\frac{-\gamma}{1+\gamma}}]\label{53},
\end{eqnarray}
where $y(x)=e^{\mu x}$. The EoS for combined fluid is
\begin{eqnarray}\nonumber
\omega&=&\frac{1}{(1+2m) [\frac{I(x)}{6\alpha
c^2(m+2)}+H_{c_0}^2]y(x)}\times[-(1+2m)[\frac{\mu}{(m+2)}+1]\nonumber\\
&\times&[\frac{I(x)}{6\alpha
c^2(m+2)}+H_{c_0}^2]y(x)-\frac{(1+2m)}{6\alpha
c^2(m+2)^2}[A+\rho_{c_0}B^{-(m+2)(1+\gamma)}]^{\frac{1}{1+\gamma}}\nonumber\\
&-&(1+2m)[\frac{\mu}{(m+2)}+1]H_{c_0}^2B^\mu]\label{54}.
\end{eqnarray}
\begin{figure}
\center\epsfig{file=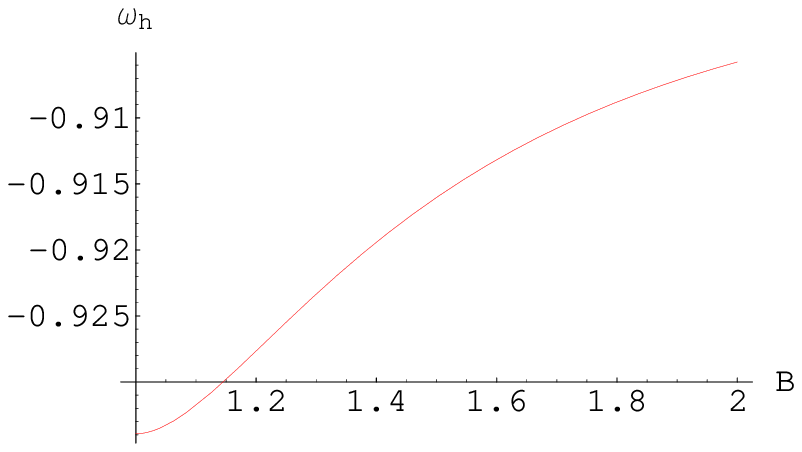, width=0.5\linewidth}\caption{Variation
of $\omega_h$ against $B$ for $\rho_{c_0}=1, ~H_{c_0}=1, ~c=2,
~\gamma=0.1, ~\alpha=0.1, ~A=1.$}
\end{figure}
\begin{figure}
\center\epsfig{file=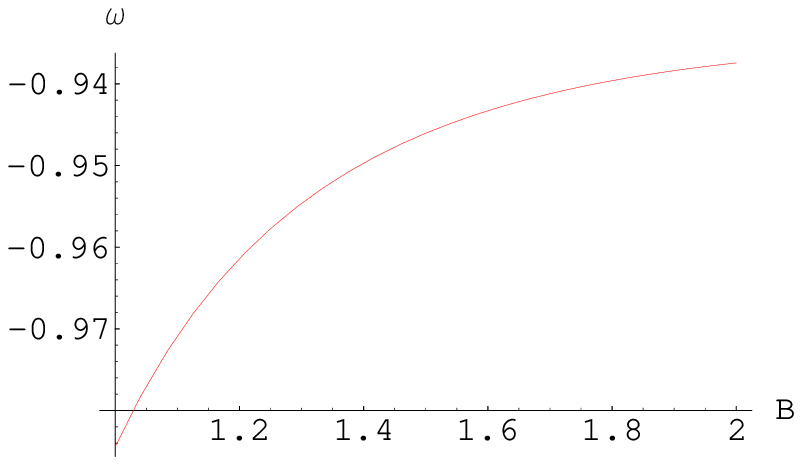, width=0.5\linewidth}\caption{Variation
of $\omega$ against $B$ for $\rho_{c_0}=1, ~H_{c_0}=1, ~c=2,
~\gamma=0.1, ~\alpha=0.1, ~A=1.$}
\end{figure}
\begin{figure}
\center\epsfig{file=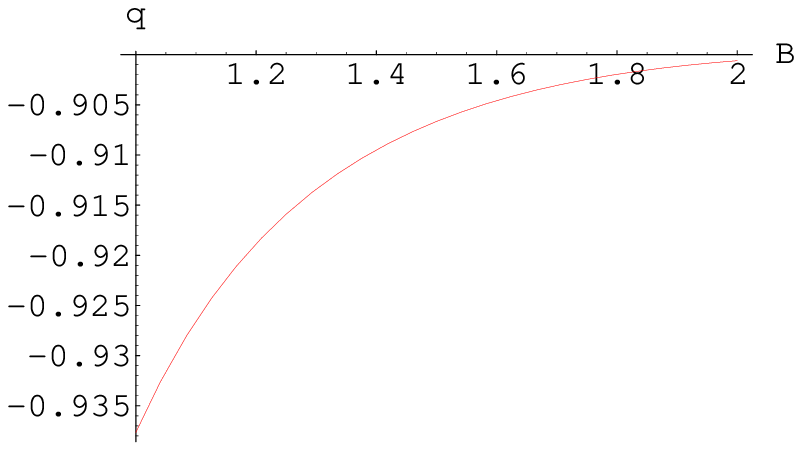, width=0.5\linewidth}\caption{Variation
of $q$ against $B$ for $\rho_{c_0}=1, ~H_{c_0}=1, ~c=2,
~\gamma=0.1, ~\alpha=0.1, ~A=1.$}
\end{figure}
\begin{figure}
\center\epsfig{file=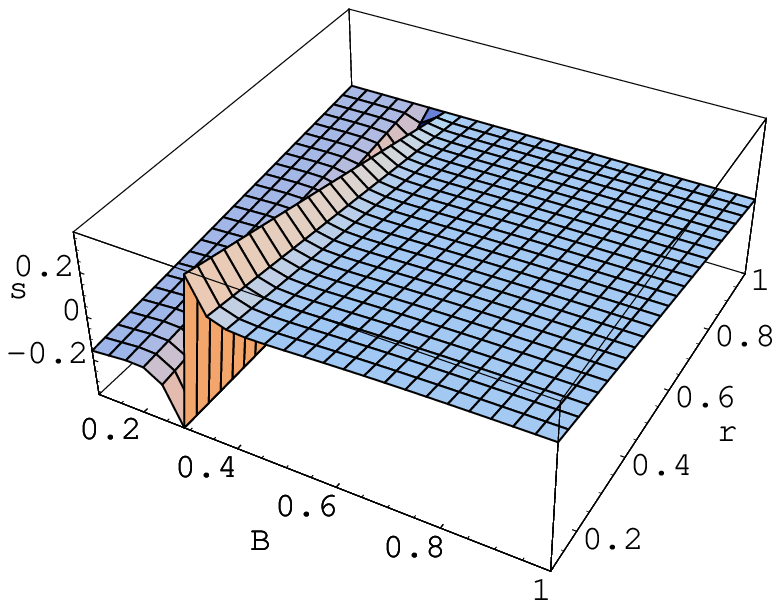, width=0.5\linewidth}\caption{Variation
of $s$ against $r$ and $B$ for $\rho_{c_0}=1, ~H_{c_0}=1, ~c=2,
~\gamma=0.1, ~\alpha=0.1, ~A=1.$}
\end{figure}
The behavior of $\omega_h$ for GHDE is shown in Figure \textbf{12}
for appropriate values which represents the expanding universe by
keeping negative sign. Figure \textbf{13} shows the variation of
$\omega$ against $B$ indicating $\omega$ as negative. The graphs of
$q$ against $B$ and $s$ against $r$ and $B$ are plotted in Figures
\textbf{14} and \textbf{15}, respectively. The deceleration
parameter also possesses negative sign and generates DE and $s$ has
increasing behavior for decreasing $r$ and increasing $B$.

\subsection{Interacting Case}

Now we assume that the universe is a combination of GHDE and GCG
interacting each other. The equations of conservation for
interacting GHDE and GCG become
\begin{eqnarray}\label{55}
\dot{\rho}_c+(m+2)H_2(\rho_c+p_c)&=&-(m+2)\delta
H_2\rho_c,\\\label{56}
\dot{\rho}_h+(m+2)H_2(\rho_h+p_h)&=&(m+2)\delta H_2\rho_c.
\end{eqnarray}
Using EoS of GCG in Eq.(\ref{56}) and after solving, we obtain
\begin{equation}\label{57}
\rho_c=[\frac{A}{1+\delta}+\rho_{c_1}B^{-(m+2)(1+\gamma)(1+\delta)}]
^{\frac{1}{1+\gamma}},
\end{equation}
where $\rho_{c1}$ is the constant of integration. From
Eq.(\ref{42}), we obtain
\begin{equation}\label{58}
p_c=-A[\frac{A}{1+\delta}+\rho_{c_1}B^{-(m+2)(1+\gamma)(1+\delta)}]^
{\frac{-\gamma}{1+\gamma}}.
\end{equation}
Inserting $\rho_c$ in Eq.(\ref{43}), we finally obtain
\begin{eqnarray}\nonumber
H_2^{2}&=&\frac{1}{6\alpha c^2(m+2)}e^{\mu x}
\int[\frac{A}{1+\delta}+\rho_{c_1}e^{-(m+2)(1+\gamma)(1+\delta)x}]
^{\frac{1}{1+\gamma}}e^{-\mu x}dx\nonumber\\
&+&H_{c_1}^2B^\mu,\label{60}
\end{eqnarray}
where $H_{c_1}$ is another constant of integration. Thus we have
\begin{eqnarray}\label{61}
\rho_h&=&\frac{(1+2m)}{6\alpha c^2(m+2)}e^{\mu x}J(x)-
[\frac{A}{1+\delta}+\rho_{c_1}B^{-(m+2)(1+\gamma)(1+\delta)}]
^{\frac{1}{1+\gamma}}\nonumber\\&+&(1+2m)H_{c_1}^2B^\mu,
\end{eqnarray}
where
\begin{equation*}
J(x)=\int[\frac{A}{1+\delta}+\rho_{c_1}e^{-(m+2)(1+\gamma)(1+\delta)x}]^
{\frac{1}{1+\gamma}}e^{-\mu x}dx.
\end{equation*}
The energy conservation equation gives
\begin{equation}\label{62}
p_h=\delta\rho_c-\rho_h-\frac{1}{(m+2)}\frac{\partial\rho_h}{\partial
x},
\end{equation}
\begin{figure}
\center\epsfig{file=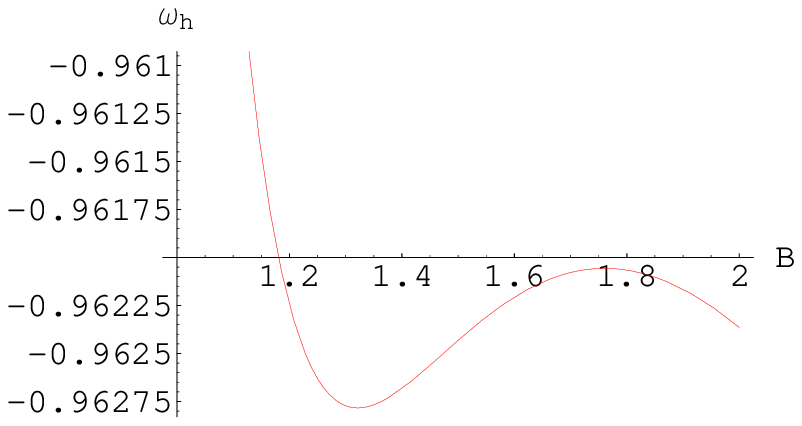, width=0.5\linewidth}\caption{Variation of
$\omega_h$ against $B$ for $\rho_{c_1}=1,~H_{c_1}=1,~c=2,
~\gamma=0.1,~\alpha=0.1,~A=1,~\delta=0.01.$}
\end{figure}
\begin{figure}
\center\epsfig{file=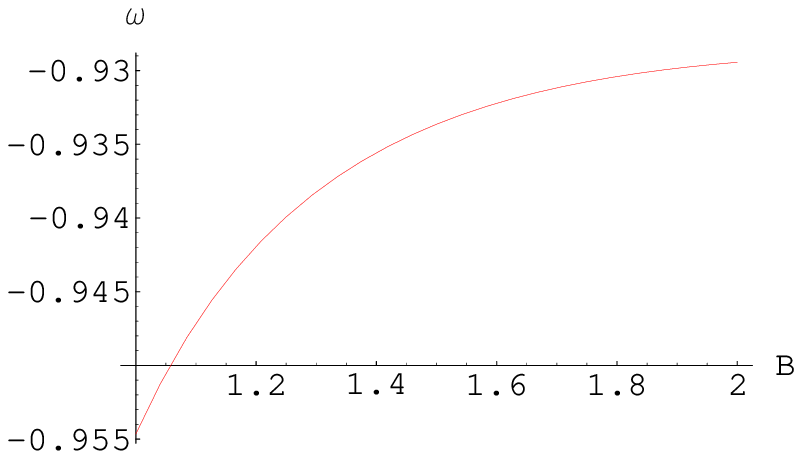, width=0.5\linewidth}\caption{Variation
of $\omega$ against $B$ for $\rho_{c_1}=1, ~H_{c_1}=1, ~c=2,
~\gamma=0.1, ~\alpha=0.1, ~A=1, ~\delta=0.01.$}
\end{figure}
\begin{figure}
\center\epsfig{file=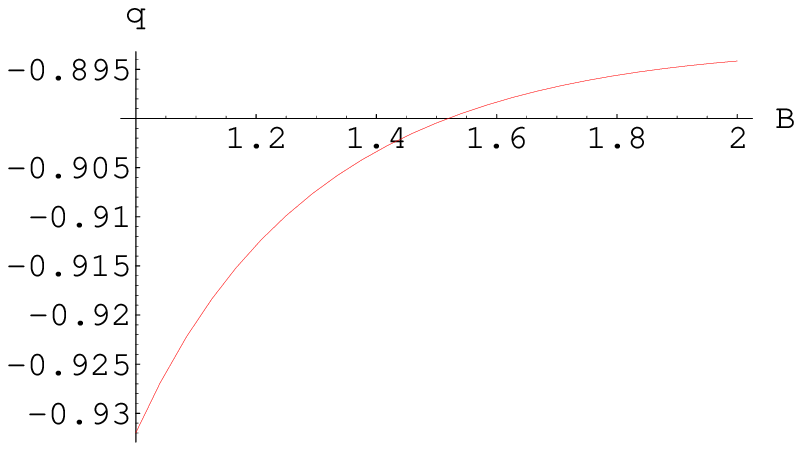, width=0.5\linewidth}\caption{Variation
of $q$ against $B$ for $\rho_{c_1}=1, ~H_{c1}=1, ~c=2,
~\gamma=0.1, ~\alpha=0.1, ~A=1, ~\delta=0.01.$}
\end{figure}
\begin{figure}
\center\epsfig{file=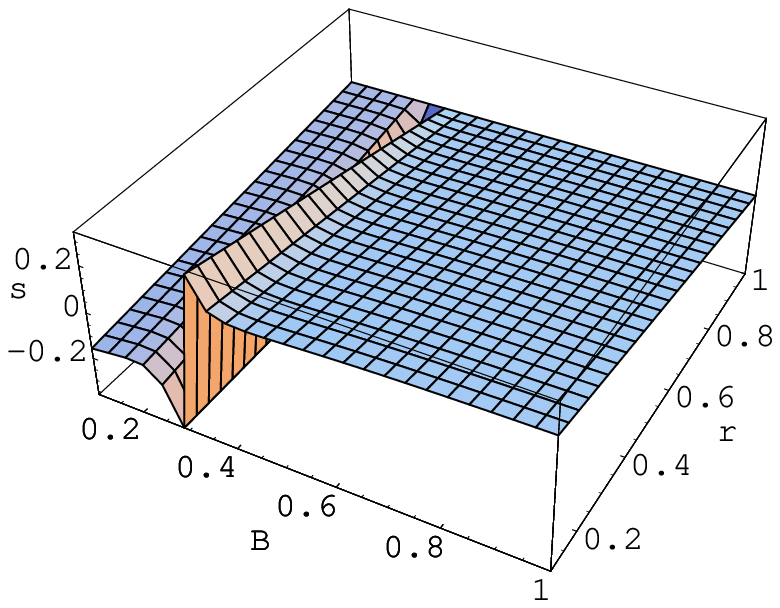, width=0.5\linewidth}\caption{Variation
of $s$ against $r$ and $B$ for $\rho_{c_1}=1, ~H_{c1}=1, ~c=2,
~\gamma=0.1, ~\alpha=0.1, ~A=1, ~\delta=0.01.$}
\end{figure}
Using Eq.(\ref{57}) and (\ref{61}), it follows that
\begin{eqnarray}\label{63}
p_h&=&-(1+2m)[\frac{\mu}{(m+2)}+1][\frac{J(x)}{6\alpha
c^2(m+2)}-H_{c_1}^2]y(x)\nonumber\\&+&[\delta-\frac{(1+2m)}{6\alpha
c^2(m+2)}][\frac{A}{1+\delta}+\rho_{c1}B^{-(m+2)(1+\gamma)(1+\delta)}]
^{\frac{1}{1+\gamma}}\nonumber\\&+&[\frac{A}{1+\delta}-\delta\rho_{c_1}
B^{-(m+2)(1+\gamma)(1+\delta)}]\nonumber\\&\times&[\frac{A}{1+\delta}+\rho_{c1}B^{-(m+2)
(1+\gamma)(1+\delta)}]^{\frac{-\gamma}{1+\gamma}} ,
\end{eqnarray}
where $y(x)=e^{\mu x}$.

The EoS for GHDE $\omega_h$ is expressed as
\begin{eqnarray}\nonumber
\omega_h&=&\frac{1}{(1+2m)[\frac{\mu}
{(m+2)}+1][\frac{J(x)}{6\alpha c^2(m+2)}-H_{c_1}^2]y(x)-
[\frac{A}{1+\delta}+\rho_{c_1}B^{-(m+2)(1+\gamma)(1+\delta)}]^
{\frac{1}{1+\gamma}}}\nonumber\\&\times&-(1+2m)[\frac{\mu}{(m+2)}+1]
[\frac{J(x)}{6\alpha
c^2(m+2)}-H_{c_1}^2]y(x)\nonumber\\&+&[\delta-\frac{(1+2m)}{6\alpha
c^2(m+2)}][\frac{A}{1+\delta}+\rho_{c_1}B^{-(m+2)(1+\gamma)(1+\delta)}]
^{\frac{1}{1+\gamma}}\nonumber\\&+&[\frac{B}{1+\delta}-\delta\rho_{c_1}
B^{-(m+2)(1+\gamma)(1+\delta)}][\frac{B}{1+\delta}+\rho_{c_1}B^
{-(m+2)(1+\gamma)(1+\delta)}]^{\frac{-\gamma}{1+\gamma}}\label{64}.
\end{eqnarray}
The EoS for interacting GHDE and GCG has the form
\begin{eqnarray}\nonumber
\omega&=&\frac{1}{(1+2m) [\frac{\mu}{(m+2)}+1][\frac{J(x)}{6\alpha
c^2(m+2)}-H_{c_1}^2]y(x)-[\frac{A}{1+\delta}+\rho_{c_1}B^{-(m+2)
(1+\gamma)(1+\delta)}]^{\frac{1}{1+\gamma}}}\\\nonumber&\times&-
(1+2m)[\frac{\mu}{(m+2)}+1][\frac{J(x)}{6\alpha
c^2(m+2)}+H_{c_1}^2]y(x)\label{65}\\&-&\frac{(1+2m)}{6\alpha
c^2(m+2)^2}[\frac{A}{1+\delta}+\rho_{c_1}B^{-(m+2)(1+\gamma)
(1+\delta)}]^{\frac{1}{1+\gamma}}.
\end{eqnarray}
All the parameters are shown in Figures \textbf{(16)-(19)}. These
plots indicate that this model also generates DE for interacting
case.

\section{Concluding Remarks}

The main purpose of this paper is to check the role of EoS,
deceleration, and statefinder parameters in the accelerated
universe. To this end, we have considered the LRS BI universe model
with two models of DE, i.e., GHDE/GRDE. We have assumed different
cases like combined DM and DE fluid, GHDE/GRDE without and with DM
(both interacting and non-interacting) as well as GHDE/GRDE without
and with GCG (both interacting and non-interacting) to investigate
evolution of the universe. In all these cases, the deceleration, EoS
and statefinder parameters are calculated which help to explore the
accelerating universe.

We have displayed all the parameters graphically by using
appropriate values of the constants to understand their behavior.
The EoS and deceleration parameters contain negative sign for both
interacting and non-interacting scenarios which support the
acceleration of the universe. The graphical behavior of statefinder
parameters shows that $s$ always increases as $r$ decreases and $B$
increases for specified values. The general nature of the parameters
indicate the accelerating behavior of the universe. We would like to
mention here that the nature of GRDE model can be obtained by
replacing $\alpha$ with $1-\beta$ in GHDE model. We have also
calculated the values of $\alpha$ and $\beta$ for accelerating
universe. It has been found that our universe will be accelerating
if $\alpha<\frac{(m+2)^2c^2-3(1+2m)}{(7m^2-8m+10)c^2}$ for GHDE and
$\beta>\frac{(6m^2-12m+6)c^2+3(1+2m)}{(7m^2-8m+10)c^2}$ for GRDE. We
conclude that statefinder parameters have an extra contribution of
EoS parameter $\omega$ as compared to FRW. The statefinder
diagnostic pair also represents the $\Lambda CDM$ model.

Finally, we would like to mention here that our work supports the
results of a recent paper \cite{22} for FRW model, where all the
parameters represent DE and evolution of the universe.


\begin{thebibliography}{40}

\bibitem{1} Perlmutter, S. et al.: Astron. Soc.
\textbf{29}(1997)1351; Nature \textbf{391}(1998)51; Astrophys. J.
\textbf{517}(1999)565; Riess, A.G. et al.: Astron. J.
\textbf{116}(1998)1009.

\bibitem{2} Spergel, D.N. et al.: Astrophys. J. Suppl. \textbf{170}(2007)377.

\bibitem{3} Tegmark, M. et al.: Phys. Rev. \textbf{D69}(2004)03501.

\bibitem{4} Bennett, C.L. et al.: Astrophys. J. Suppl. \textbf{148}(2003)1;
Peebles, P.J.E. and Ratra, B.: Rev. Mod. Phys. \textbf{75}(2003)559.

\bibitem{5} Ratra, B. and  Peebles, P.J.E.: Phys. Rev. \textbf{D37}(1998)3406;
Padmanabhan, T.: Gen. Relativ. Gravit. \textbf{40}(2008)529.

\bibitem{6} Chiba, T., Okabe, T. and Yamaguchi, M.: Phys. Rev. \textbf{D62}(2000)023511.

\bibitem{7} Sen, A.: JHEP \textbf{48}(2002)204; Padmanabhan, T.: Phys. Rev.
\textbf{D66}(2002)021301.

\bibitem{8} Caldwell, R.R.: Phys. Lett. \textbf{B545}(2002)23; Nojiri, S. and Odintsov, S.D.:
Phys. Lett. \textbf{B562}(2003)147; ibid. \textbf{B565}(2003)1.

\bibitem{9} Bento, M.C., Bertolami, O. and Sen, A.A.: Phys. Rev.
\textbf{D66}(2002)043507; Liu, D.J. and Li, X.Z.: Chin. Phys.
Lett. \textbf{22}(2005)1600.

\bibitem{10} Susskind, L.: J. Math. Phys. \textbf{36}(1995)6377.

\bibitem{11} Cohen, A., Kaplan, D. and Nelson, A.: Phys. Rev. Lett. \textbf{82}(1999)4971.

\bibitem{12} Hsu, S.D.H.: Phys. Lett. \textbf{B594}(2004)13; Li, M.: Phys. Lett.
\textbf{B603}(2004)1.

\bibitem{13} Zhang, X.: Int. J. Mod. Phys. \textbf{D14}(2005)1597.

\bibitem{14} Li, H., Guo, Z.K. and Zhang, Y.Z.:
Int. J. Mod. Phys. \textbf{D15}(2006)869; Almeida, J.P.B. and
Pereira, J.G.: Phys. Lett. \textbf{B636}(2006)75; Gong, Y.: Phys.
Rev. \textbf{D70}(2004)064029.

\bibitem{15} Gao, C., Wu, F., Chen, X. and Shen, Y. G.: Phys. Rev. \textbf{D79}(2009)043511.

\bibitem{16} Chattopadhyay, S.: Europ. Phys. J. Plus \textbf{127}(2012)16.

\bibitem{17} Feng, C.J. and Li, X.Z.: Phys. Lett. \textbf{B680}(2009)355.

\bibitem{18} Cai, R.G., Hu, B. and Zhang, Y.: Commun. Theor. Phys. \textbf{51}(2009)954.

\bibitem{18+} Kim, K.Y., Lee, H.W., Myung, Y.S.: Gen. Relativ. Gravit. \textbf{43}(2011)1095.

\bibitem{19} Sadjadi, H.M. and Vadood, N.:
JCAP \textbf{0808}(2008)036; Saridakis, E.N.: Phys. Lett.
\textbf{B661}(2008)335; Granda, I.N. and Oliveros, A.: Phys. Lett.
\textbf{B669}(2008)275; Feng, C.J. Phys. Lett.
\textbf{B672}(2009)94; Jamil, M., Farooq, M.U. and Rashid, M.A.:
Eur. Phys. J. \textbf{C61}(2009)471.

\bibitem{21} Xu, L., Lu, J. and Li, W.: Eur. Phys. J. \textbf{C64}(2009)89.

\bibitem{22} Khatua, P.B. and Debnath. U.: arXiv:1106.5689.

\bibitem{23} Sharif, M. and Zubair, M.: Astrophys. Space Sci.
\textbf{330}(2010)399; Sharif, M. and Waheed, S.: Eur. Phys. J.
\textbf{C72}(2012)1876.

\bibitem{24} Sahni, V., Saini, T., Starobinsky, A.A., and Alam, U.: JETP Lett. \textbf{77}(2003)201.

\bibitem{25} Setare, M. R.: Phys. Lett. \textbf{B648}(2007)329.

\bibitem{26} Gorini, V., Kamenshchik, A. and Moschella, U.: Phys. Rev.
\textbf{D67}(2003)063509; Alam, U., Sahni, V., Saini, T.D., and
Starobinsky, A.A.: Mon. Not. Roy. Astron. Soc.
\textbf{344}(2003)1057.

\end{thebibliography}
\end{document}